\documentclass[twocolumn,nofootinbib,amsmath,amssymb,a4paper]{revtex4}

\usepackage{graphicx}
\usepackage{dcolumn}
\usepackage{bm}
\usepackage{relsize}
\usepackage{pstcol}

\RequirePackage{xspace}

\providecommand{\skz}{\mbox{$S$}}
\providecommand{\ckz}{\mbox{$C$}}

\newcommand{\timesix}{\ensuremath{\times10^{6}}}

\newcommand{\DE}{\ensuremath{\Delta E}}
\newcommand{\mb}{\ensuremath{m_{\rm ES}}}

\providecommand{\dt}{\deltat}
\newcommand{\ttag}{\ensuremath{t_{\rm tag}}}

\newcommand\etal{{\it et al.}}

\newcommand{\msp}{\ensuremath{\phantom{-}}}

\newcommand{\bfig}{\begin{figure}[htbpc!]}
\newcommand{\efig}{\end{figure}}
\newcommand\bef{\begin{figure}}
\newcommand\edf{\end{figure}}
\newcommand\dbline{\noalign{\vskip 0.10truecm\hrule}\noalign{\vskip 2pt}\noalign{\hrule\vskip 0.10truecm}}

\newcommand\sgline{\noalign{\vskip 0.10truecm\hrule\vskip 0.10truecm}}
\newcommand\beq{\begin{equation}}
\newcommand\eeq{\end{equation}}
\newcommand\bear{\begin{array}}
\newcommand\enar{\end{array}}
\newcommand\beqa{\begin{eqnarray}}
\newcommand\eeqa{\end{eqnarray}}
\newcommand\ben{\begin{enumerate}}
\newcommand\een{\end{enumerate}}

\newcommand{\etagg}{\ensuremath{\eta_{\gaga}}}
\newcommand{\etappp}{\ensuremath{\eta_{3\pi}}}

\newcommand{\etaprg}{\ensuremath{\etapr_{\rho\gamma}}}

\newcommand{\etapeppgg}{\ensuremath{\etapr_{\eta(\gamma\gamma)\pi\pi}}}
\newcommand{\etapeppppp}{\ensuremath{\etapr_{\eta(3\pi)\pi\pi}}}

   \newcommand{\rhoz}{\ensuremath{\rho^0}}

\newcommand{\fetapiz}{\ensuremath{\eta\piz}\xspace}
\newcommand{\etapiz}{\ensuremath{\Bz\to\fetapiz}\xspace}

\newcommand{\fetapKz}{\ensuremath{\etapr K^0}}
\newcommand{\fetapKs}{\ensuremath{\etapr\KS}}
\newcommand{\fetapKl}{\ensuremath{\etapr\KL}}
\newcommand{\etapKz}{\ensuremath{\Bz\to\fetapKz}}
\newcommand{\etapKs}{\ensuremath{\Bz\to\fetapKs}}
\newcommand{\etapKl}{\ensuremath{\Bz\to\fetapKl}}

\newcommand{\fetappiz}{\ensuremath{\etapr\piz}\xspace}
\newcommand{\etappiz}{\ensuremath{\Bz\to\fetappiz}\xspace}

\newcommand{\retappiz}{\ensuremath{xx^{+xx}_{-xx} \pm xx}\xspace}

\def\babar{\mbox{\slshape B\kern-0.1em{\smaller A}\kern-0.1em
    B\kern-0.1em{\smaller A\kern-0.2em R}}}
\def\CP                {\ensuremath{C\!P}\xspace}
\def\Bz      {\ensuremath{B^0}\xspace}
\def\Bbar    {\kern 0.18em\overline{\kern -0.18em B}{}\xspace}
\def\Bzb     {\ensuremath{\Bbar^0}\xspace}
\def\BB      {\ensuremath{B\Bbar}\xspace} 
\newcommand{\etapr}{\ensuremath{\eta^{\prime}}\xspace}
\def\Kz    {\ensuremath{K^0}\xspace}
\def\KS    {\ensuremath{K^0_{\scriptscriptstyle S}}\xspace} 
\def\KL    {\ensuremath{K^0_{\scriptscriptstyle L}}\xspace} 
\def\piz   {\ensuremath{\pi^0}\xspace}
\def\gaga  {\ensuremath{\gamma\gamma}\xspace}
\def\qqbar {\ensuremath{q\overline q}\xspace}
\def\mes        {\mbox{$m_{\rm ES}$}\xspace}
\def\deltat{\ensuremath{{\rm \Delta}t}\xspace}
\def\deltamd{\ensuremath{{\rm \Delta}m_d}\xspace}
\def\pipi  {\ensuremath{\pi^+\pi^-}\xspace}
\def\twpiz  {\ensuremath{\pi^0\pi^0}\xspace}
\def\stwob{\ensuremath{\sin\! 2 \beta   }\xspace}
\def\invfb   {\ensuremath{\mbox{\,fb}^{-1}}\xspace}
\providecommand{\tcp}{\mbox{$t_{\CP}$}}
\def\DS{\ensuremath{{\rm \Delta}S}\xspace}

\newcommand{\jprlBase}       {Phys.\ Rev.\ Lett.\xspace}
\newcommand{\jprBase}        {Phys.\ Rev.\xspace}
\newcommand{\jplBase}        {Phys.\ Lett.\xspace}
\newcommand{\npBase}         {Nucl.\ Phys.\xspace}
\newcommand{\jhepBase}       {J.~High Energy Phys.\xspace}
\newcommand{\plb}       [1]  {\jplBase\ B~{\bf #1}}
\newcommand{\jprl}      [1]  {\jprlBase\ {\bf #1}}
\newcommand{\jprd}      [1]  {\jprBase\ D~{\bf #1}}

\newcommand{\npb}       [1]  {\npBase\ B~{\bf #1}}
\newcommand{\jhep}      [1]  {\jhepBase\ {\bf #1}}
\newcommand{\rpizpizBA}{\ensuremath{1.48\pm0.26\pm0.12}}
\newcommand{\rpizpizBe}{\ensuremath{1.1\pm0.3\pm0.1}}
\renewcommand{\retappiz}{\ensuremath{2.8\pm1.0\pm0.3}}
\newcommand{\fpizpiz}{\ensuremath{\piz\piz}}
\renewcommand{\fetapiz}{\ensuremath{\eta\piz}}
\renewcommand{\fetappiz}{\ensuremath{\etapr\piz}}
\newcommand{\fetaeta}{\ensuremath{\eta\eta}}
\newcommand{\fetaetap}{\ensuremath{\eta\etapr}}
\newcommand{\fetapetap}{\ensuremath{\etapr\etapr}}
\newcommand{\pizpiz}{\ensuremath{\Bz\to\fpizpiz}}
\renewcommand{\etapiz}{\ensuremath{\Bz\to\fetapiz}}
\renewcommand{\etappiz}{\ensuremath{\Bz\to\fetappiz}}
\newcommand{\etaeta}{\ensuremath{\Bz\to\fetaeta}}
\newcommand{\etaetap}{\ensuremath{\Bz\to\fetaetap}}
\newcommand{\etapetap}{\ensuremath{\Bz\to\fetapetap}}

\begin{document}

\title{CP Violation in \etapKz\ and Status of SU(3)-related Decays}

\author{J.~G. Smith}
 \email{jgsmith@pizero.colorado.edu}
\affiliation{%
Physics Department, University of Colorado, Boulder, CO 80309-0390
}%

\begin{abstract}
We present measurements from Belle and \babar\ of the time-dependent 
\CP-violation parameters \skz\ and \ckz\ in \etapKz\ decays.  Both
experiments observe mixing-induced \CP violation with a significance of 
more than 5 standard deviations in this $b\to s$ penguin dominated mode.  
We also compare with theoretical expectations and discuss the latest
results for SU(3)-related decays which are useful for obtaining bounds
on the expected values of \skz\ and \ckz.
\end{abstract}

\maketitle

\section{Introduction}

Because of the large rate for the process, the decay \etapKz\ has proved
to be the most precise outside of the $c\bar c\Kz$ system for
determination of the value of \stwob\ through time-dependent
\CP-violation measurements.  The dominant process is a
penguin (loop) decay where new physics can enter through additional
particles in the loop.  There have been many predictions for the
Standard-Model (SM) and non-SM expectations for this and related
processes \cite{theory,BN,WZ,GLNQ,GRZ}.

We report new results from \babar\ \cite{babar} and Belle \cite{belle}.
Both of these analyses observe mixing-induced \CP violation with a
significance of more than 5 standard deviations.  We compare the results
with theoretical expectations, some of which use data (summarized below)
for decays of \Bz\ mesons to pairs of isoscalar mesons.

\section{Status of \etapKz}

When the decay \etapKz\ was first observed \cite{CLEOetapobs}, the 
measured branching was much larger than theoretical predictions
involving naive factorization.  The situation has changed substantially 
with more recent calculations.  In QCD factorization calculations \cite{BN},
it was pointed out that higher-order QCD corrections and slight tweaking
of parameters can easily account for the large observed result, though
theory errors are still large.  In a 
paper involving QCD factorization, SCET and inputs from $B$-decay data,
the explanation of the large rate is thought to come from
``charming-penguins", long-distance effects involving the $c\bar c$ in
the loop.  While the details of the explanation for the large rate still
differ somewhat, the recent calculations account for the large observed
branching fraction with predominantly penguin amplitudes and the
contribution from tree or penguin amplitudes involving $V_{ub}$
is small.  This feature is important as will be seen below.

\section{Experimental details}

For these measurements, Belle uses a dataset with a luminosity of 492
\invfb\ (535\timesix\ \BB\ pairs).  The corresponding numbers for
\babar\ are 349 \invfb\ and 384\timesix\ \BB\ pairs.  

Both experiments use five final states of \etapKs, denoted $B_{\CP}$. Those 
with a $\KS\to\pipi$ decay use the decays $\etapr\to\rhoz\gamma$
(\etaprg), $\etapr\to\etagg\pipi$ (\etapeppgg), and
$\etapr\to\etappp\pipi$ (\etapeppppp), where \etagg\ and \etappp\
denote the decays $\eta\to\gaga$ and $\eta\to\pipi\piz$, respectively.
Those with a $\KS\to\piz\piz$ use \etaprg\ and \etapeppgg.
For the decays \etapKl, both experiments use the \etapeppgg\ mode while
Belle additionally uses \etapeppppp.

The quantities used in the analysis are similar for both experiments: a
$B$-mass variable (denoted \mes\ for \babar), $\DE\equiv E_B^*-E_{\rm
beam}^{*}$ (the asterisk denotes center-of-mass quantities), variables
that discriminate between spherical $B$-decay events and jetlike \qqbar\
background, a tagging variable to determine the flavor of the ``tag" $B$
($B_{\rm tag}$), and the difference $\deltat \equiv \tcp - \ttag$ of the
proper decay times $\tcp$ and $\ttag$ of the \CP\ and tag $B$ mesons, 
respectively.  Maximum-likelihood (ML) fits are used to distinguish signal
from background and to determine the parameters $S$ and $C$ via the time
dependence (for \babar---the formula for Belle is similar)
\begin{eqnarray}
  F(\dt) &=&
        \frac{e^{-\left|\deltat\right|/\tau}}{4\tau} [1 \mp\Delta w \pm
                                                   \label{eq:FCPdef}\\
   &&\hspace{-1em}(1-2w)\left(-\xi S\sin(\deltamd\deltat) -
C\cos(\deltamd\deltat)\right)]\,\nonumber
\end{eqnarray}
where $\xi$ is the \CP\ eigenvalue of the final state ($-1$ for \etapKs, 
$+1$ for \etapKl).
The upper (lower) sign denotes a decay accompanied by a \Bz (\Bzb) tag,
$\tau$ is the mean $\Bz$ lifetime, $\deltamd$ is the mixing frequency,
and the mistag parameters $w$ and $\Delta w$ are the average and
difference, respectively, of the probabilities that a true $\Bz$\ is
incorrectly tagged as a $\Bzb$\ or vice versa.  

For \etapKs\ the ML fits for \babar\ use all of the inputs mentioned above;
Belle uses all except the event shape (they cut on this quantity).  The
fits for \etapKl\ are similar except that one of the pair [\mes,\DE] is not 
used in the fit.  Instead both experiments perform a constrained fit to the 
$B$-decay hypothesis since the \KL\ energy is poorly measured.

\begin{figure}[!htb]
\hspace*{-0.5cm}
 \includegraphics[angle=0,width=0.5\textwidth]{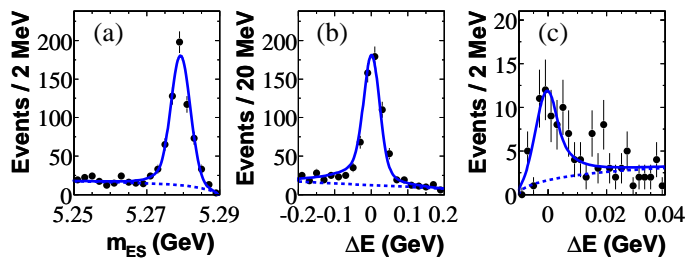}\\
\caption{\label{fig:projMbDE}
Distributions of \babar\ data projected 
onto (a) \mb\ and (b) \DE\ for \etapKs\ candidates, and (c) \DE\ for
\etapKl\ candidates.  The solid lines shows
the full fit result and the dashed lines show the background
contributions.}
\end{figure} 

\begin{table*}[!ht]
\caption{Summary for \babar\ and Belle of the number of events entering the 
fits ($N_{\rm fit}$), signal yields ($N_{\rm sig}$), and fit values of $S$ and $C$ 
for each subsample.  Belle reports $A=-C$.} 
\label{tab:results}
\vspace*{-0.5cm}
\begin{center}
\begin{tabular}{lrcccrccc}
\dbline
     & \multicolumn{4}{c}{\babar}&\multicolumn{4}{c}{Belle}\\
Mode &$N_{\rm fit}$&$N_{\rm sig}$&$-\xi S$&$C$&$N_{\rm fit}$&$N_{\rm sig}$&$-\xi S$&$C$\\
\sgline
$\etaprg K^0_{\pipi}$	&11943&$566\pm30$&$0.56\pm0.14$    &$-0.24\pm0.10$
                        & 2870&$794\pm36$&$\msp0.59\pm0.15$&$\msp0.14\pm0.10$\\
$\etapeppgg K^0_{\pipi}$&664  &$224\pm16$&$0.61\pm0.23$    &$-0.26\pm0.14$
                        &634 &$363\pm21$&$\msp0.94\pm0.22$ &$-0.08\pm0.13$\\
$\etapeppppp K^0_{\pipi}$&177 &$73\pm9$  &$0.89\pm0.35$    &$\msp0.14\pm0.25$
                         &125 &$100\pm11$&$\msp0.78\pm0.47$&$-0.12\pm0.27$\\
$\etaprg K^0_{\twpiz}$ &13915&$133\pm24$ &$0.56\pm0.41$    &$\msp0.15\pm0.27$
                        &683  &$103\pm15$&$-0.04\pm0.38$   &$-0.32\pm0.28$\\
$\etapeppgg K^0_{\twpiz}$&490 &$52\pm9$  &$0.84\pm0.42$    &$-0.26\pm0.36$
                         &247 &$62\pm9$  &$\msp1.27\pm0.35$&$\msp0.17\pm0.38$\\
\sgline
$\etapKs$                &    &          &$0.62\pm0.11$    &$-0.18\pm0.07$
                         &    &          &$\msp0.67\pm0.11$&$\msp0.03\pm0.07$\\
\sgline
$\etapeppgg\KL$	       &4199 &$204\pm24$ &$0.32\pm0.28$    &$\msp0.08\pm0.23$
                       &4606 &$393\pm17$ &$\msp0.40\pm0.25$&$-0.09\pm0.17$\\
$\etapeppppp\KL$       &$-$   &$-$       &$-$              & $-$
                       &585   &$ 62\pm13$&$\msp0.86\pm0.73$&$-0.07\pm0.44$\\
\sgline
$\fetapKz$	       &      & &$0.58\pm0.10\pm0.03$&$-0.16\pm0.07\pm0.03$
                       &      & &$\msp0.64\pm0.10\pm0.04$&$\msp0.01\pm0.07\pm0.05$\\
\dbline
\end{tabular}
\end{center}
\end{table*}

\section{Results from Belle and \babar}

\begin{figure}[!htb]
  \begin{center}
   \includegraphics[width=0.23\textwidth]{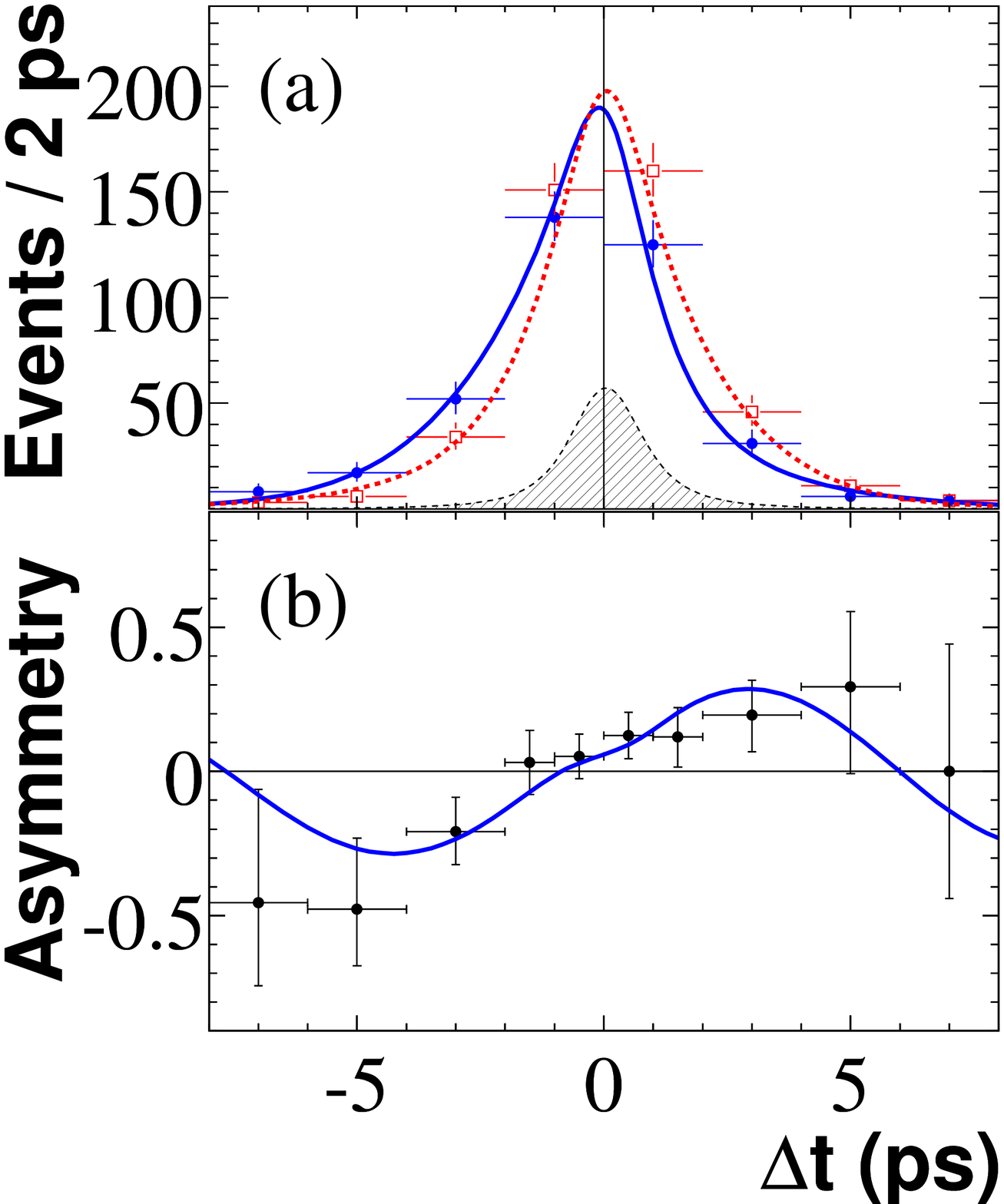}
   G\includegraphics[width=0.23\textwidth]{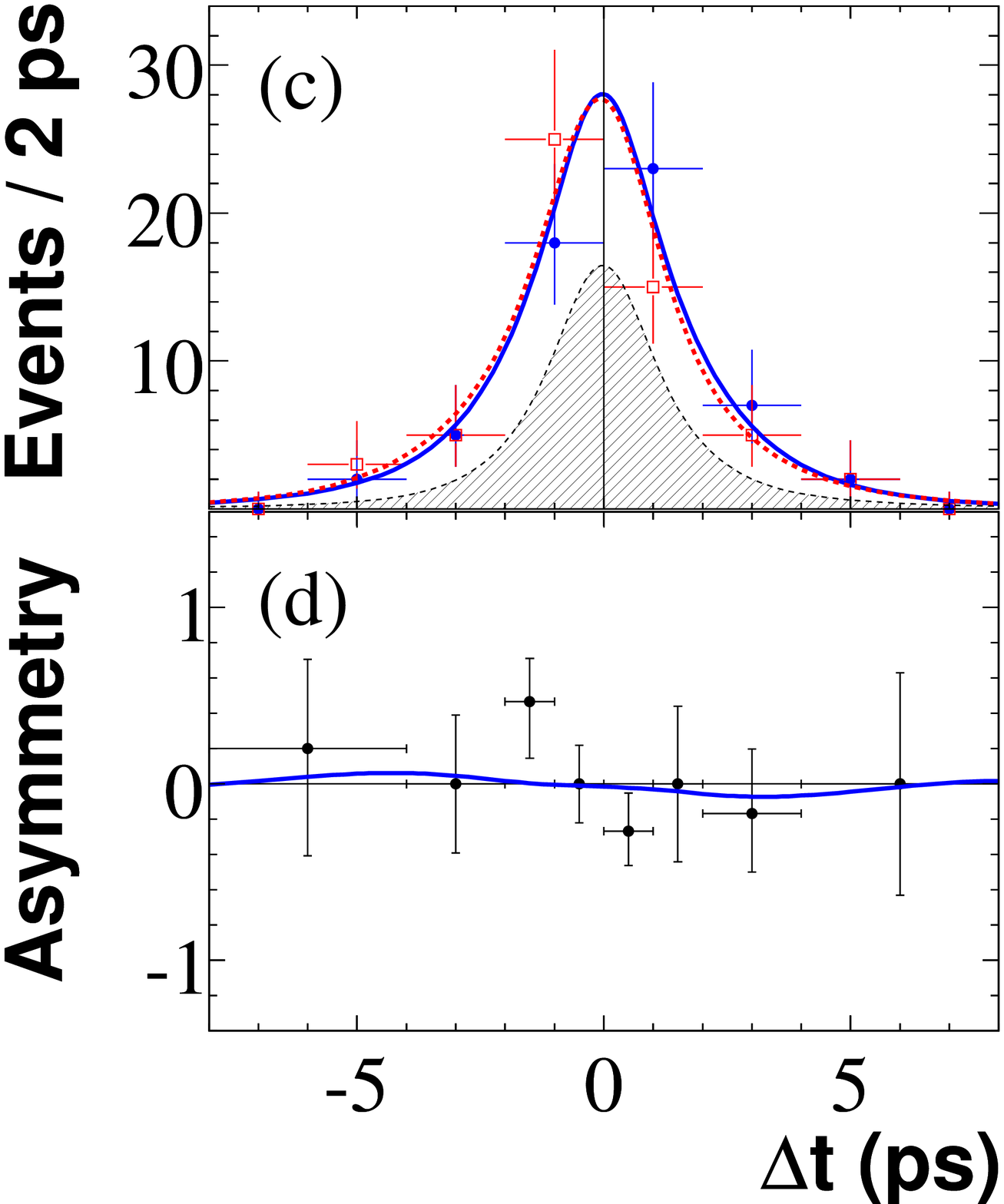}
\end{center}
  \vspace*{-0.5cm}
 \caption{\babar\ projections
onto \dt\ for (a) \etapKs\ and (c) \etapKl\ of the data (points 
with error bars for \Bz\ tags  in red empty rectangles and \Bzb\ tags 
in blue solid circles), fit function (red dashed and blue solid
lines for \Bz\ and \Bzb\ tagged events, respectively), and background
function (black shaded regions). We show the asymmetry between \Bz\ and \Bzb\ 
tags for (b) \etapKs\ and (d) \etapKl; the lines represent the fit functions.}
\label{fig:babar_dt}
\end{figure}

\begin{figure}[!htb]
  \begin{center}
   \includegraphics[width=0.45\textwidth]{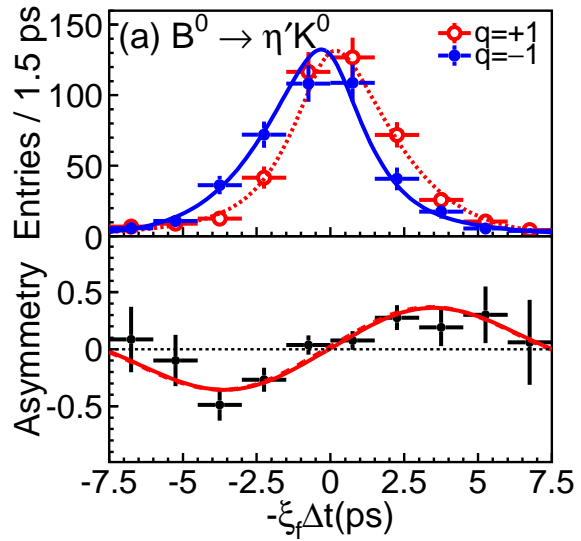}
\end{center}
  \vspace*{-0.2cm}
\caption{Top: Belle background-subtracted \dt\ distributions for \etapKz\ data 
(points with error bars for \Bz\ tags in red empty circles and \Bzb\ tags
in blue solid circles), fit function (red dashed and blue solid
lines for \Bz\ and \Bzb\ tagged events, respectively).
Bottom: Asymmetry between \Bz\ and \Bzb\ tags, where the line represents
the fit function.}
  \label{fig:belle_dt}
\end{figure}

To illustrate the samples, we show in Fig.~\ref{fig:projMbDE} projection
plots for the signals from \babar.  The results of the fits for both
experiments for all subsamples are shown in Table \ref{tab:results}.
The values of $S$ are inconsistent with zero at the level of 5.5
standard deviations ($\sigma$) for \babar\ and 5.6$\sigma$ for Belle.
$C$ is consistent with zero for both experiments.  Plots of the
time-dependence for \babar\ are shown in Fig.~\ref{fig:babar_dt} and Belle
in Fig.~\ref{fig:belle_dt}.  The values of $S$ and $C$ for both
experiments are now in good agreement with average values
$S=0.61\pm0.07$ and $C=-0.09\pm0.06$ \cite{HFAG}.

\begin{table*}[t!]
\caption{Experimental measurements for the decays used as input to the
GLNQ and GRZ calculations along with the coefficients that multiply the 
branching fractions for each mode and the 90\% CL UL used for the
recent GRZ calculation.}
\label{tab:isoscalar}
\begin{center}
\begin{tabular}{lcccccc}
\dbline
Mode& GLNQ Coeff.&GRZ Coeff.&\multicolumn{4}{c}{BR or 90\% CL ULs $(10^{-6})$}\\
    &            &          & BABAR & Belle & CLEO & GRZ UL\\
\sgline
\red{\etaetap} &\red{0.96}&\red{0.87}&\red{$<1.7$}\cite{BABARetapeta}&\red{$<4.0$}\cite{Belleetapeta}&\red{$<27$}\cite{CLEOetapeta}&\red{$<1.7$}\\
\red{\etappiz}&\red{0.59}&\red{0.23}&\red{$<2.1$}\cite{BABARetapeta}&\red{\retappiz}\cite{Belleetappiz}&\red{$<5.7$}\cite{CLEOetappiz}&\red{$<2.4$}\\
\etapetap &0.53&&$<2.4$\cite{BABARetapetap}&$<7.7$\cite{Belleetapeta}&$<47$\cite{CLEOetapeta} &$<2.4$\\
\etaeta &0.38&&$<1.8$\cite{BABARetapetap}&$<2.0$\cite{Belleetaeta}&$<18$\cite{CLEOetapeta}&$<1.8$\\
\red{\etapiz} &\red{0.33}&\red{0.83}&\red{$<1.3$}\cite{BABARetapeta}&\red{$<2.5$}\cite{Belleetaeta}&\red{$<2.9$}\cite{CLEOetappiz}&\red{$<1.3$}\\
\pizpiz &0.14&&\rpizpizBA\cite{BABARpizpiz}&\rpizpizBe\cite{Bellepizpiz}&$<4.4$\cite{CLEOpizpiz}&$<1.58$\\
\dbline
\end{tabular}
\end{center}
\end{table*}

\section{Theoretical understanding}

\subsection{First-principles calculations}

The theoretical expectation is that the value of $C$ (indicative of
direct \CP\ violation) is near zero while for the SM, $S$ should be
nearly equal to \stwob.  The world average for \stwob\ is
$0.675\pm0.026$ \cite{HFAG}.
Small deviations from this value arise from tree or penguin $b\to u\bar u s$
amplitudes which have a different weak phase.  The size of these
deviations is expected to be $\sim$$\pm0.01$; when calculation parameters
are varied, the range is $\pm0.03$ \cite{theory}.

\subsection{SU(3)-related modes and theoretical limits}

The above predictions are based on QCD factorization, PQCD, or SCET 
calculations.  In 2003, Grossman, Ligeti, Nir, and Quinn (GLNQ) \cite{GLNQ}
showed that SU(3) and data from related $B$ decays can be used to limit
the size of the $b\to u\bar u s$ amplitudes.  In their analysis, data
from the six processes \etaetap, \etappiz, \etapetap, \etaeta, \etapiz,
and \pizpiz\ are used.  Measurements for these decays have improved
substantially in recent years, with the current experimental situation
summarized in Table \ref{tab:isoscalar}.  

\begin{figure}[!hb]
  \begin{center}
   \includegraphics[width=0.45\textwidth]{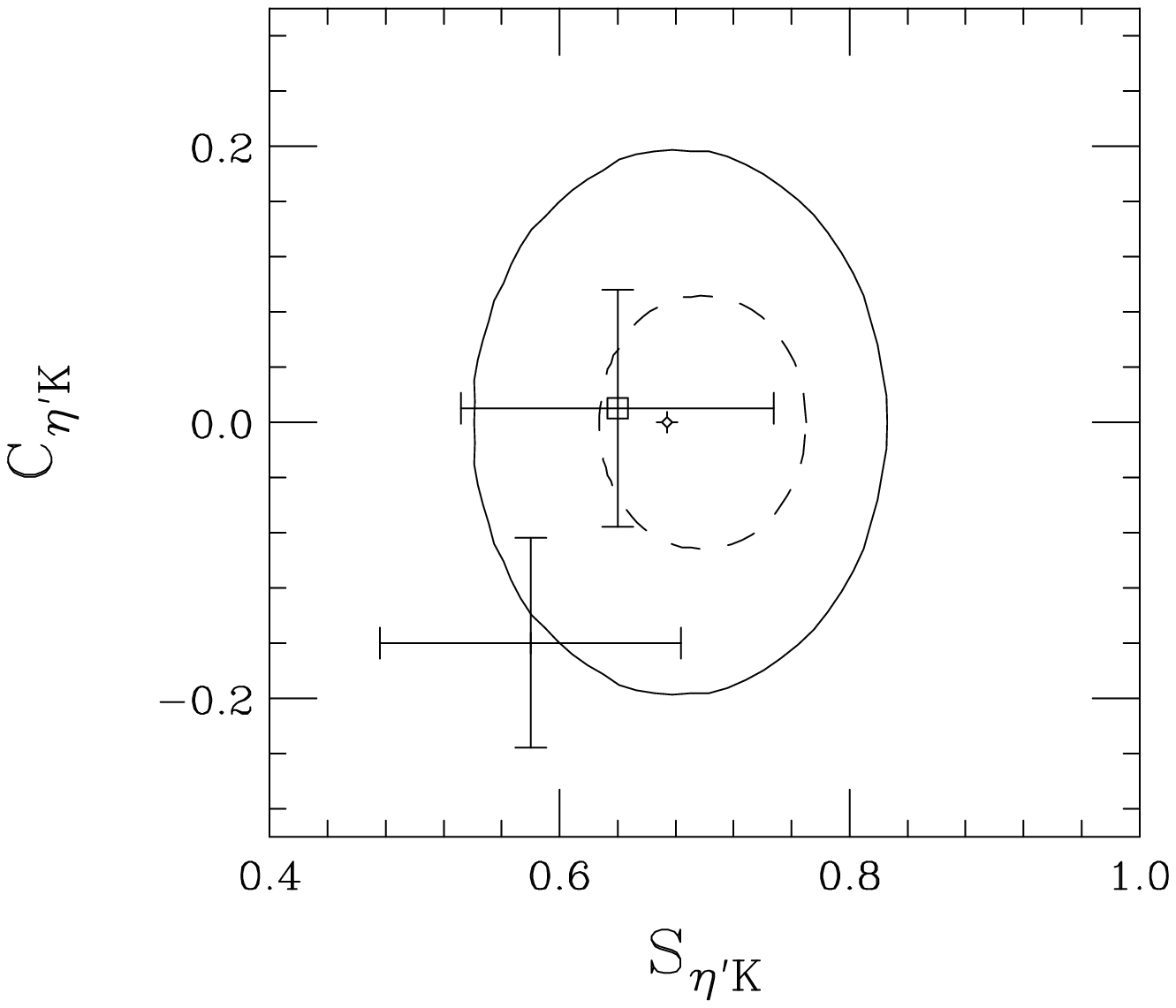}
\end{center}
  \vspace*{-0.5cm}
 \caption{The curves indicate the allowed ranges for $S$ and $C$ for
\etapKz.  The solid curve gives the bounds with the full GLNQ analysis
with six inputs and the dashed line gives the more restrictive case
where exchange and penguin-annihilation amplitudes are neglected so that
only \etapiz, \etappiz, and \etaetap\ enter.  The point (square) with
error bars shows the \babar\ (Belle) measured result.  The small plotted
point near the center shows the average of $c\bar{c}\Kz$ measurements.}
  \label{fig:GRZ}
\end{figure}

In this table, we also show the coefficients from the GLNQ calculation
and order the modes from largest to smallest contributors to the limit
on \DS, deviation in the value of $S$ from \stwob.  We also show in red
the three decays that are used in a more recent update of the GLNQ
calculation by Gronau, Rosner, and Zupan (GRZ) \cite{GRZ}, where it is 
assumed that exchange and penguin annihilation
amplitudes are small.  They find the allowed ranges of $S$ and $C$ shown in 
Fig.~\ref{fig:GRZ}.  The range from the \etaetap, \etappiz, and \etapiz\
decays is $-0.046<\DS<0.094$.

\section{Summary}
The \babar\ and Belle experiments have each measured \CP\ violation in
the \etapKz\ decay with significance greater than 5$\sigma$.  Belle
finds $S=0.64\pm0.10\pm0.04$ and $C=0.01\pm0.07\pm0.05$ while \babar\
finds $S=0.58\pm0.10\pm0.03$ and $C=-0.16\pm0.07\pm0.03$.  These
measurements are in good agreement with the expectations of the Standard
Model.  While the precision of these measurements is much better than
was anticipated a decade ago, substantial improvement in
precision is needed in order to check for non-SM effects.

\begin{acknowledgments}
I would like to thank the organizers for an enjoyable and productive meeting.  
It is also important to acknowledge all of the people at KEK-B and PEP-II for 
their superb efforts in producing luminosities beyond what we expected a
decade ago.  I also wish to thank colleagues on \babar\ and Belle for
their efforts in achieving the very impressive results presented here.
\end{acknowledgments}

\end{document}